\documentclass[aps,floatfix,showpacs,twocolumn]{revtex4}
\usepackage{graphicx,bm}
\begin{document}
\title {One particle in a box: the simplest model for a Fermi gas in the unitary limit}
\author{Ludovic Pricoupenko$^{1}$ and Yvan Castin$^{2}$}
\affiliation
{$^{1}$Laboratoire de Physique Th\'{e}orique des Liquides, Universit\'{e} Pierre et Marie Curie,
case courier 121, 4 place Jussieu, 75252 Paris Cedex 05, France. \\
$^{2}$Laboratoire Kastler Brossel, Ecole normale sup\'erieure, 24 rue Lhomond, 75231 Paris Cedex 05, France.}
\date{\today}
\begin{abstract}
We consider a single quantum particle in a spherical box interacting 
with a fixed scatterer at the center, to construct a model of a degenerate
atomic Fermi gas close to a Feshbach resonance. One of the key predictions of 
the model is the existence of two branches for the macroscopic state of the 
gas, as a function of the magnetic field controlling the value of the 
scattering length. This model is able to draw a qualitative picture of all 
the different features recently observed in a degenerate atomic Fermi gas 
close to the resonance, even in the unitary limit.  

\end{abstract}
\maketitle
Recently several experiments have produced two spin-component
degenerate Fermi gases in the unitary limit, that is in a limit where
the scattering length $a$ of two atoms with different spin components
is much larger than the mean interparticle separation
\cite{Thomas1,Thomas2,Jin,Salomon1,Salomon2}. Similar experiments have also 
been performed with bosonic atoms \cite{Wieman,Grimm,Durr,Ketterle1}. This was 
made possible by a tuning of the scattering length virtually from $-\infty$ 
to $+\infty$ using a Feshbach resonance driven by an external uniform 
magnetic field, as first demonstrated on bosonic atoms 
\cite{Ketterle2,Cornell}.

This regime constitutes a theoretical challenge. Some theories, based on 
the Hartree-Fock mean field approximation for the normal
gas or on the BCS mean field calculation for the superfluid phase, rely on 
the small parameter $k_F a$, where the Fermi momentum of the 
gas is conventionally related to the mean total density $\rho$ by the 
non-interacting case formula:
\begin{equation}
k_F = \left( 3\pi^2\rho\right)^{1/3} \quad .
\label{eq:kF}
\end{equation}
Such approaches are not quantitative in the limit $k_F |a|\to +\infty$.
More sophisticated approaches have been developed to extend
the accuracy of the theory to the unitary regime 
\cite{NSR,Randeria,Heiselberg,Combescot,Panda}.
The difficulty comes from the fact that there exists no obvious small
parameter for the theory in the unitary limit, at least in the degenerate
regime of a temperature $T$ much smaller than the Fermi temperature $T_F$, 
which is the regime
considered here and also the present experimental situation.
Note that, in the classical regime $T\gg T_F$, one recovers a small parameter 
$\rho^{1/3} |f| \ll 1$ since the typical scattering amplitude $f$ for 
$|a| \to +\infty$, is on the order of $\lambda$, where $\lambda$ is the
thermal de Broglie wavelength \cite{note1,note2}.

The scope of the present work is to give the simplest possible physical picture
of a two spin-component Fermi gas for arbitrary values of the scattering length
$a$, essentially at zero temperature. The model is not made to give 
quantitative predictions, but, as we shall see, it qualitatively reproduces 
the experimental observations, and it will provide, we hope, useful 
guidelines for experiments to come.

\begin{figure}[htb]
\resizebox{8 cm}{!}{\includegraphics{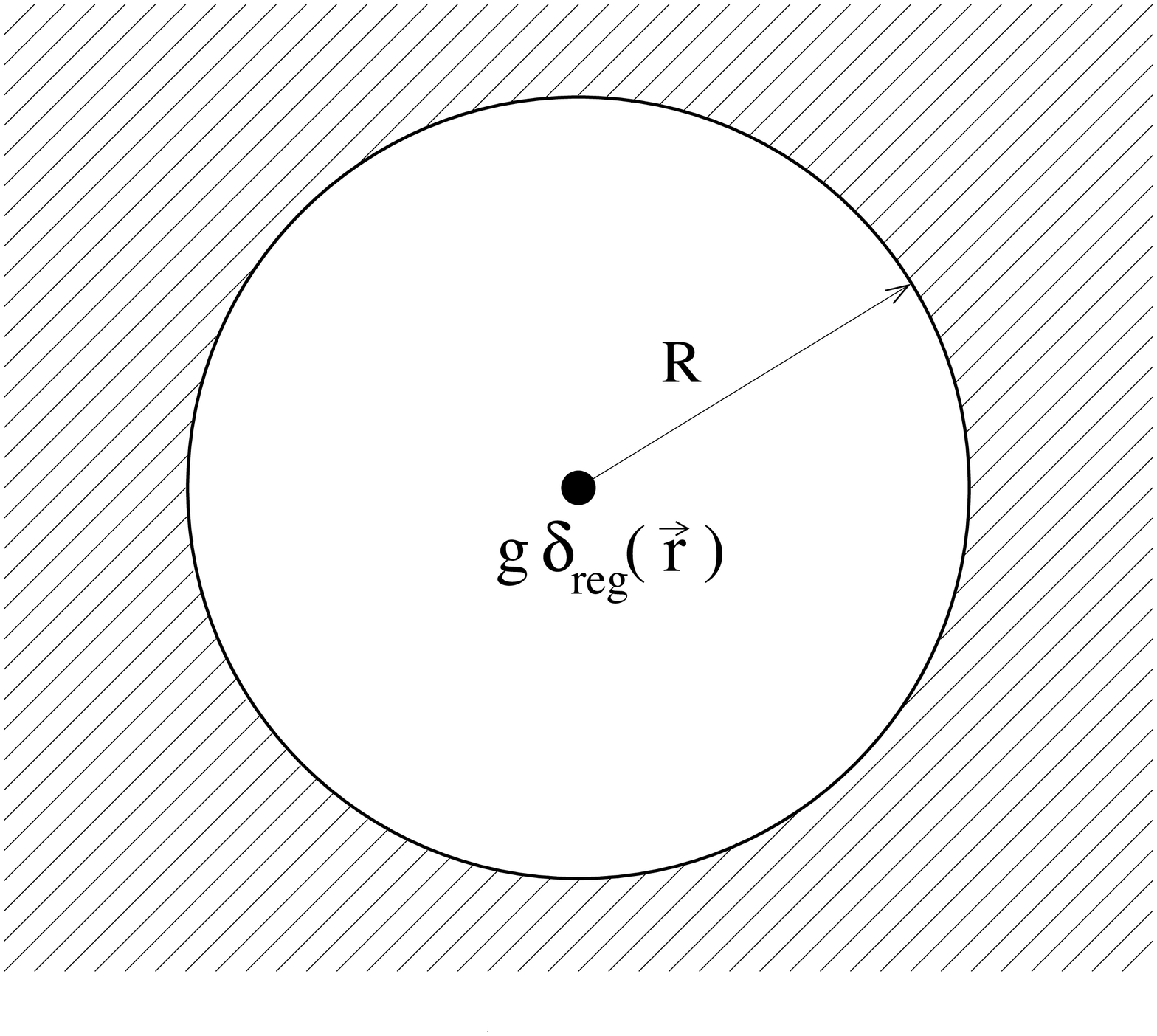}}
\caption{Model used: a fictitious particle corresponding to the relative motion
of a fermion with spin $+1/2$ and the nearest fermion with spin $-1/2$ is 
scattered by the fixed delta Fermi pseudo-potential \cite{Castin} in the center
of a spherical box of radius $R$ with absorbing boundary conditions. The box
mimics the interaction effect of the $N/2-1$ remaining spin $-1/2$ atoms and 
the Fermi statistical effect of the remaining $N/2-1$ spin $+1/2$ fermions. 
The coupling constant $g$ is $4\pi\hbar^2a/m$, where $a$ is the scattering
length of two fermions with opposite spin, $g \delta_{\rm reg}(\vec{r}\,) = g
\delta(\vec{r}\,) \partial_r (r.)$ is the Fermi 
pseudo-potential which imposes the contact condition
(\ref{eq:contact}) on the two-body wave function \cite{Cohen,Castin,Olshanii,Petrov}.}
\label{fig:model}
\end{figure}

\noindent{\bf The model:}
We consider a spatially homogeneous gas of $N/2$ fermions of spin $+1/2$ and 
$N/2$ fermions of spin $-1/2$, each particle having a mass $m$. The 
interaction of a given spin $+1/2$ particle with the $N/2$ spin $-1/2$ 
particles is modelized (i) by the interaction of a fictitious
particle of mass equal to the reduced mass $m/2$, with a fixed scatterer at 
the center of a spherical box of radius $R$, and (ii) with the boundary 
conditions that the wavefunction $\phi(\vec{r}\,)$ of the fictitious 
particle vanishes on the surface of the box. (i): The interaction of the 
fictitious particle with the scatterer represents the interaction
of the given spin $+1/2$ particle with its nearest spin $-1/2$ neighbour: in 
the model, these two opposite spin particles are considered in their center 
of mass frame and in the singlet spin state. (ii): The boundary condition 
mimics the interaction effect of the $N/2-1$ other spin $-1/2$ particles and 
the Fermi statistical effect of the remaining $N/2-1$ spin $+1/2$ particles, 
see Figure~\ref{fig:model}. A similar model was very recently put forward for 
bosons, with the difference that the box is replaced by a harmonic potential 
\cite{Greene}.

In the absence of interactions between the fermions, the total energy of the 
gas is given by the known ideal Fermi gas formula in the thermodynamic limit:
\begin{equation}
E = \frac{3}{5} N  \epsilon_F \quad , 
\label{eq:ideal}
\end{equation}
where the Fermi energy $\epsilon_F= \hbar^{2} k_F^{2}/2m$ is related to the mean 
density through (\ref{eq:kF}). The gas energy $E$ is also related to the 
energy $\epsilon$ of the fictitious particle in the spherical box by
\begin{equation}
E = \frac{1}{2} N \epsilon \quad .
\label{eq:ener_tot}
\end{equation}
For the ideal Fermi gas, there is no scatterer in the box so that the ground 
state value of $\epsilon$ is
\begin{equation}
\epsilon_0 = \frac{\hbar^{2}}{m} \left(\frac{\pi}{R}\right)^{2} \quad .
\end{equation}
This establishes the link between the radius $R$ and the mean density
of the gas \cite{theotherway}~:
\begin{equation}
k_F R = \left(\frac{5}{3}\right)^{1/2}\pi \quad .
\label{eq:kFR}
\end{equation}

In presence of interactions, the scatterer in the model is the Fermi delta 
pseudo-potential with a coupling constant $g=4 \pi\hbar^2a /m$ where $a$ is 
the $s$-wave scattering length of two fermions with opposite spin components 
\cite{Cohen,Castin,Olshanii,Petrov}. The choice of such a zero-range potential is 
allowed in the regime 
\begin{equation}
\rho r_e^{3} \ll 1 \quad ,
\label{eq:nal}
\end{equation}
where $r_e$ is the effective range of the true interaction potential,
and which allows to consider these systems
as gases rather than liquids even in the unitary limit.
The two-body scattering amplitude derived from  the pseudo-potential,
\begin{equation}
                          f_k =-\frac{1}{ a^{-1} + ik  }
\end{equation}
is indeed 
a good approximation of the scattering amplitude in simple models of a 
Feshbach resonance:
\begin{equation}
                         f_k^{\rm Fesh} = -\frac{1}{a^{-1} + ik  - k^2 r_e/2},
\end{equation}
when $k_F | r_e | \ll  1$ since $k$ is at most on the order of $k_F$
at $T\ll T_F$. $r_e$ was calculated in \cite{Kokkelmans} and one finds that 
condition (\ref{eq:nal}) is very well satisfied in present experiments on 
Li${}^6$ \cite{Combescot}.

We will use the fact that the pseudo-potential is equivalent to replacing
the interaction potential by the contact condition \cite{Olshanii,Petrov,Cohen}
\begin{equation}
\lim_{r\to 0} \frac{\partial_r(r\phi)}{r\phi} = -\frac{1}{a} \quad ,
\label{eq:contact}
\end{equation}
where $r$ is the distance to the origin. Out of the origin in the box,
the wavefunction then
solves the free Schr\"odinger equation:
\begin{equation}
-\frac{\hbar^{2}}{m}\Delta \phi = \epsilon \phi \quad .
\end{equation}
Restricting to the $s$-wave, which is the only partial wave in which the 
pseudo-potential scatters, $\phi$ is rotationally symmetric and we set 
$\phi(r)=u(r)/r$. For positive energies $\epsilon=\hbar^{2} k^{2}/m$, where 
$k>0$, one then finds
\begin{equation}
u(r) \propto \sin [k(r-R)] \quad ,
\end{equation}
and the contact condition Eq.(\ref{eq:contact}) imposes:
\begin{equation}
\tan kR = k a \quad .
\end{equation}
For negative energies $\epsilon=-\hbar^{2}\kappa^{2}/m$, where $\kappa>0$, one 
finds
\begin{equation}
u(r) \propto \sinh[k(r-R)] \quad ,
\end{equation}
where Eq.(\ref{eq:contact}) imposes:
\begin{equation}
\tanh \kappa R = \kappa a \quad .
\end{equation}

\begin{figure}[htb]

\resizebox{8 cm}{!}{\includegraphics{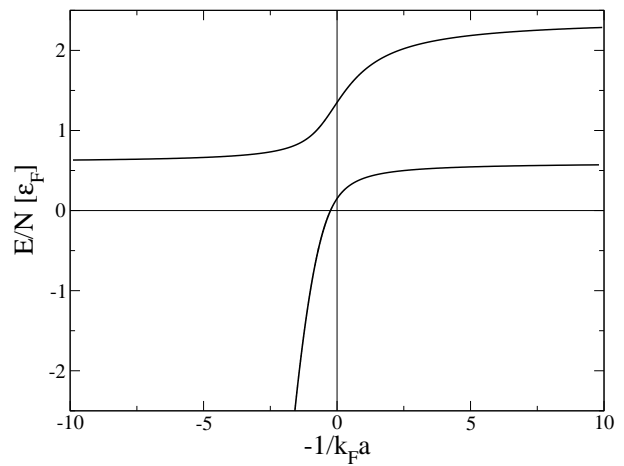}}

\caption{Total energy of the gas in units of the Fermi energy, as function
of $-1/k_F a$. Only the first two branches of the model are represented.}
\label{fig:ener}
\end{figure}

\noindent {\bf Discussion:}
The energy spectrum of the previous model allows to calculate the energy of 
the gas according to Eq.(\ref{eq:ener_tot}). The result is shown in Figure 
\ref{fig:ener} as function of $-1/k_F a$, restricting to the two lowest energy 
branches. The choice of the parameter $-1/k_F a$ is inspired by the way the 
experimental results on $^6$Li are usually presented: in this way, the right 
part of the Figure \ref{fig:ener} corresponds to magnetic fields
above the Feshbach resonance, and the left part to magnetic fields below 
the resonance. The existence of several energy branches for the macroscopic 
state of the gas is a first crucial prediction of the model. The ground 
branch connects two clearly identified regimes:
\begin{itemize}
\item the first regime $k_F a\to 0^{-}$ (on the right hand side of Figure 2) is 
a weakly attractive Fermi gas which corresponds at zero temperature to the 
BCS phase in a more complete treatment.
\item the second regime $k_F a\to 0^{+}$ (on the left hand side of
Figure 2) corresponds to a dilute gas of dimers;
each dimer corresponds to the bound state of the two-body problem in free
space since the size of the box $R$, on the order of the mean interatomic 
distance, is here much larger than the spatial extension $\sim a$
of the dimer. The energy of a dimer is $-\hbar^{2}/m a^{2}$ which explains
the quadratic drop of the ground energy on the left part of the resonance.
In a more complete treatment, one would find at $T=0$ that these dimers 
form a Bose-Einstein condensate since they are bosons in the limit 
$\rho^{1/3} a \ll 1$.
\end{itemize}
The upper branch in Figure \ref{fig:ener} is also easy to identify in the 
weakly interacting limit $k_F a \to 0^{+}$, where it corresponds to a weakly
repulsive Fermi gas. It clearly corresponds to a metastable state of the gas:
a three body collision between atoms will form a dimer plus an extra atom 
carrying away the binding energy, a mechanism that will depopulate the upper 
branch and populate the ground branch. This constitutes a first experimental 
way to produce dimers from an atomic gas, already proposed in the case of
bosons in \cite{Pricou}, 
with the disadvantage that the resulting molecules have a high
center of mass kinetic energy, on the order of $\hbar^{2}/m a^{2}$.
These strong three-body losses prevent in present experiments to follow
adiabatically the upper branch in Figure \ref{fig:ener} from the left
part of the Figure ($a>0$) to the right part of the Figure ($a<0$).
In the unitary regime, our model predicts in the upper branch an energy
per particle scaling as $\hbar^{2}\rho^{2/3}/m$ larger than the ideal Fermi
gas case, that is corresponding to an effective repulsion.
A similar behavior was predicted for bosons by A. Leggett \cite{Leggett}.

\begin{figure}[htb]
\resizebox{8 cm}{!}{\includegraphics{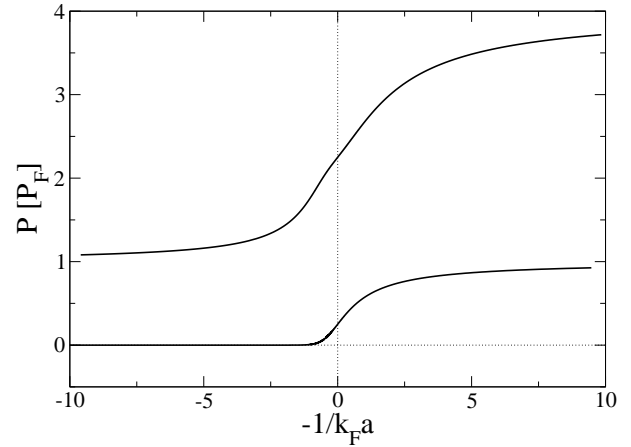}}
\caption{Pressure $P$ of the gas in units of the Fermi pressure 
$P_F=\hbar^2k_F^5/15\pi^2m$, as function of $-1/k_F a$. Only the first two
branches of the model are represented.}
\label{fig:pressure}
\end{figure}

Inspection of Figure \ref{fig:ener} immediately inspires another way
to produce a gas of ultracold molecules than the one relying on three-body
inelastic collisions: one starts with a weakly attractive atomic Fermi gas, 
on the right of the Feshbach resonance, and one slowly crosses
the resonance from right to left in order to follow adiabatically the 
ground branch. Note that a similar mechanism has been developed in the 
case of an atomic Bose gas \cite{Mies}.

Another important feature of the model is that nothing dramatic
happens right on the Feshbach resonance, the mean energy per particle
being simply proportional to the Fermi energy with some numerical factor 
(here 3/40). As this mean energy is less than the ideal gas value 
(\ref{eq:ideal}), the gas experiences an effective attraction due to the 
atomic interactions. This universality and this effective attraction
appeared already in several approaches \cite{Heiselberg,Combescot,note1,note2},
and are confirmed by the experimental results \cite{Thomas2,Salomon1}.
As the total energy remains positive, so will be the pressure and the 
compressibility of the gas at the Feshbach resonance: the Fermi gas will not 
experience a collapse when $k_F a$ becomes large in absolute value and negative,
in agreement with the recent experiments
\cite{Thomas1,Thomas2,Jin,Salomon1,Salomon2} 
and contrarily to what was feared a few years ago based on the calculation of
the compressibility in the mean field approximation \cite{Houbiers}. 

To check in a systematic way the stability of the gas in our model, we have
calculated the pressure (see Figure \ref{fig:pressure}) and the compressibility
(not shown) of the gas for the lowest two branches for an arbitrary value
of $a$. The compressibility and the pressure are positive everywhere.
Moreover, the pressure on the ground branch is lower than the ideal Fermi gas
pressure and tends exponentially to zero from above in the molecular limit 
$a\to 0^{+}$, revealing the absence of a non-zero scattering length of dimers 
in our model \cite{Dima}. The decrease of the pressure across the Feshbach 
resonance from the $a<0$ part to the $a>0$ part is observable in a trap 
through a shrinking of the size of the cloud.

In conclusion, the heuristic model that we have presented contains all the 
essential features of a two spin-component Fermi gas for $s$-wave interactions
with an arbitrary scattering length, such as the existence of a metastable 
atomic phase (upper branch) and a molecular phase (lower branch) for 
$a\to 0^+$ and the continous connection between the molecular regime
and the weakly attractive regime when $a$ is varied from $0^+$ to $0^-$ across
the Feshbach resonance.

{\bf Note:} after the submission of this paper, both scenarios for the 
production of a molecular condensate have been realized experimentally:
the adiabatic scenario \cite{Jin2,Christophe} and the scenario assisted
by three-body inelastic collisions \cite{Grimm2,Ketterle}.

We thank A. Leggett, C. Lobo, A. Minguzzi, V.R. Pandharipande,
C. Salomon and his group for useful discussions on the subject. 
Laboratoire de Physique Th\'eorique des Liquides is the Unit\'e Mixte
de Recherche 7600 of Centre National de la Recherche Scientifique.
Laboratoire Kastler Brossel is a research unit of \'Ecole normale
sup\'erieure and of Universit\'e Pierre et Marie Curie, associated to 
Centre National de la Recherche Scientifique.


\begin{thebibliography}{99}

\bibitem{Thomas1}
K. M. O'Hara, S. L. Hemmer, M. E. Gehm,
S.~R.~Granade, and J. E. Thomas, Science {\bf 298}, 2179 (2002). 

\bibitem{Thomas2} 
M. E. Gehm, S. L. Hemmer, S. R. Granade,
K.~M.~O'Hara, J. E. Thomas, Phys. Rev. A 68, 011401(R) (2003).

\bibitem{Jin} 
C. A. Regal, C. Ticknor, J. L. Bohn, and D.~S.~Jin, 
Nature {\bf 424}, 47 (2003).

\bibitem{Salomon1} 
T. Bourdel, J. Cubizolles, L. Khaykovich,
K. M. F. Magalhaes, S. J. J. M. F. Kokkelmans, G.~V.~Shlyapnikov, C. Salomon,
Phys. Rev. Lett. {\bf 91}, 020402 (2003).

\bibitem{Salomon2} 
J. Cubizolles, T. Bourdel,
S. J. J. M. F. Kokkelmans, G.~V.~Shlyapnikov, C. Salomon, cond-mat/0308018.

\bibitem{Wieman}  
E. A. Donley, N. R. Claussen, S. T. Thompson,
C.~E.~Wieman, Nature {\bf 417}, 529 (2002). 

\bibitem{Grimm} 
T. Weber, J. Herbig, M. Mark, H.-C. N\"{a}gerl, and
R. Grimm, Phys. Rev. Lett. 91, 123201 (2003).

\bibitem{Durr} 
S. D\"{u}rr, T. Volz, A. Marte, G.~Rempe,
cond-mat/0307440.

\bibitem{Ketterle1} 
K. Xu, T. Mukaiyama, J. R. Abo-Shaeer, J. K. Chin,
D. E. Miller, and W. Ketterle, Phys. Rev. Lett. {\bf 91}, 210402 (2003). 

\bibitem{Ketterle2} 
S. Inouye, M. R. Andrews, J. Stenger,
H.-J. Miesner, D. M. Stamper-Kurn, and W. Ketterle, Nature {\bf 392}, 
151 (1998).

\bibitem{Cornell}  
S. L. Cornish, N. R. Claussen, J. L. Roberts,
E. A. Cornell, C.~E.~Wieman, Phys. Rev. Lett. {\bf 85} 1795 (2000).

\bibitem{NSR} P. Nozi\`eres, S. Schmitt-Rink, J. Low Temp. Phys. {\bf 59}, 
195 (1985).

\bibitem{Randeria} M. Randeria, p. 355, in {\it Bose-Einstein Condensation},
edited by A. Griffin, D. W. Snoke, S. Stringari (Cambridge University Press,
1995).

\bibitem{Heiselberg} H. Heiselberg, Phys.Rev. A {\bf 63}, 043606 (2001).

\bibitem{Combescot} R. Combescot, Phys. Rev. Lett. {\bf 91}, 120401
(2003) and   R. Combescot, New J. Phys. {\bf 5}, 86 (2003).

\bibitem{Panda} J. Carlson, S. Y. Chang, V.~R. Pandharipande,
K.~E.~Schmidt, Phys. Rev. Lett. {\bf 91}, 050401 (2003).

\bibitem{note1} 
L. Pitaevskii, oral communication during his 70th birthday 
celebration, Trento (14 March 2003).

\bibitem{note2} 
T.-L. Ho, E. Mueller, cond-mat/0306187.

\bibitem{Greene} 
B. Borca, D. Blume, C. H. Greene, cond-mat/0304341. 
\bibitem{theotherway} 
One could have also adjusted the value of the radius $R$ to reproduce
the mean energy per particle of the ground state gas in the unitary
limit $k_F |a| = + \infty$. Since this mean energy is a universal
numerical factor times the ideal Fermi gas mean energy, this changes 
Eq.(\ref{eq:kFR}) by a numerical constant on the order of unity.
Still another way for determining the radius is to 
set the value of the number of atoms in the box to two, which also 
leads to a change in the constant in Eq.(\ref{eq:kFR}).
These changes do not affect the pictures and the conclusions drawn in this 
paper since they all provide a value of $R$ on the order of $\rho^{-1/3}$.
We are therefore confident that the assumption that $R$ is independent
of $a$ gives qualitatively correct predictions whatever the value of $k_F a$.

\bibitem{Cohen}
C. Cohen-Tannoudji, Lecture Notes at Coll\`ege de France, year
1998-1999, p.IV-8.

\bibitem{Castin} 
Y. Castin, in \textit{Coherent atomic matter waves},
Lecture Notes of Les Houches Summer School, p.1-136, 
edited by R. Kaiser, C. Westbrook, and F. David, EDP Sciences and 
Springer-Verlag (2001).

\bibitem{Olshanii}
M. Olshanii, L. Pricoupenko,
Phys. Rev. Lett. {\bf 88}, 010402 (2002).

\bibitem{Petrov}
D. S. Petrov, C. Salomon, G. V. Shlyapnikov, cond-mat/0309010.

\bibitem{Kokkelmans} 
S. J. J. M. F. Kokkelmans, J. N. Milstein,
M.~L. Chiofalo, R. Walser, and M. J. Holland, Phys. Rev. A {\bf 65}, 
053617 (2002).

\bibitem{Mies} 
F. H. Mies, E. Tiesinga, and P. S. Julienne,
Phys. Rev. A {\bf 61}, 022721 (2000). 

\bibitem{Houbiers} 
M. Houbiers, R. Ferwerda, H. T. C. Stoof, W.~I.~McAlexander,
C. A. Sackett, and R. G. Hulet,  Phys. Rev. A 56, 4864 (1997).

\bibitem{Dima} 
The calculation of the dimer scattering requires indeed
an exact four-body calculation, which shows that the
molecules have a positive scattering length, see \cite{Petrov}.

\bibitem{Jin2}   
M. Greiner, C.A. Regal, D.S. Jin, Nature {\bf 426}, 537 (2003).

\bibitem{Christophe} 
T. Bourdel, L. Khaykovich, J. Cubizolles, J. Zhang, F. Chevy,
M. Teichmann, L. Tarruell, S. Kokkelmans, C. Salomon, cond-mat/0403091.

\bibitem{Grimm2}
S. Jochim, M. Bartenstein, A. Altmeyer, S. Riedl, C. Chin,     
J. H. Denschlag,  R. Grimm, Science {\bf 302}, 2101 (2003).

\bibitem{Ketterle}
M.W. Zwierlein, C.A. Stan, C.H. Schunck, S.M.F. Raupach, S. Gupta,
Z. Hadzibabic, W. Ketterle, Phys. Rev. Lett.
{\bf 91}, 250401 (2003).

\bibitem{Pricou}
Ludovic Pricoupenko, cond-mat/0006263 (2000).

\bibitem{Leggett}
A. Leggett (unpublished); G. Baym, J. Phys. B {\bf 34}, 4541 (2001).

\end{thebibliography}
\end{document}